\documentclass[conference, a4paper]{IEEEtran}
\IEEEoverridecommandlockouts
\usepackage{cite}
\usepackage{amsmath,amssymb,amsfonts}
\usepackage{algorithm}
\usepackage{algpseudocode}
\usepackage{subfig}
\usepackage{graphicx}
\usepackage{textcomp}
\usepackage{xcolor}
\usepackage{flushend}
\usepackage[square,numbers]{natbib}
\allowdisplaybreaks
\def\BibTeX{{\rm B\kern-.05em{\sc i\kern-.025em b}\kern-.08em
    T\kern-.1667em\lower.7ex\hbox{E}\kern-.125emX}}

\begin{document}
\title{COROID: A Crowdsourcing-based Companion Drones to Tackle Current and Future Pandemics}

\author{\IEEEauthorblockN{Ashish Rauniyar\IEEEauthorrefmark{1}, Desta Haileselassie Hagos\IEEEauthorrefmark{2}, Debesh Jha\IEEEauthorrefmark{3}, Jan Erik Håkegård\IEEEauthorrefmark{1}}
\IEEEauthorblockA{\IEEEauthorrefmark{1}Sustainable Communication Technologies Group, SINTEF Digital, Norway \\
\IEEEauthorrefmark{2}Division of Software and Computer Systems, KTH Royal Institute of Technology, Sweden\\
\IEEEauthorrefmark{3}Department of Radiology, Northwestern University, USA\\
Emails: \{ashish.rauniyar, jan.e.hakegard\}@sintef.no} destah@kth.se, debesh.jha@northwestern.edu}


\maketitle

\begin{abstract}
Due to the current COVID-19 virus, which has already been declared a pandemic by the World Health Organization (WHO), we are witnessing the greatest pandemic of the decade. Millions of people are being infected, resulting in thousands of deaths every day across the globe. Even the world’s best healthcare-providing countries could not handle the pandemic because of the strain of treating thousands of patients at a time. The count of infections and deaths is increasing at an alarming rate because of the spread of the virus. We believe that innovative technologies could help reduce pandemics to a certain extent until we find a definite solution from the medical field to handle and treat such pandemic situations. Technology innovation has the potential to introduce new technologies that could support people and society during these difficult times. Therefore, this paper proposes the idea of using drones as a companion to tackle current and future pandemics. Our COROID drone is based on the principle of crowdsourcing sensors data of the public's smart devices, which can correlate the reading of the infrared cameras equipped on COROID drones. To the best of our knowledge, this concept has yet to be investigated either as a concept or as a product. Therefore, we believe that the COROID drone is innovative and has a huge potential to tackle COVID-19 and future pandemics.
\end{abstract}

\vspace{0.5ex}

\section{Introduction}
The current COVID-19 virus has already been declared a pandemic by the World Health Organization (WHO)~\cite{cucinotta2020declares}. We are already witnessing the greatest pandemic of the decade. Millions of people are being infected, resulting in thousands of deaths every day across the globe~\cite{pradhan2020fear,bulut2020epidemiology}. The count of infections due to the spread of the virus is increasing at an alarming rate, at least in China, although COVID-19 vaccines have been successfully developed~\cite{jee_2022}. We believe that innovative technologies could help reduce the pandemics to a certain extent until we find a definite solution from the medical field to handle and treat such pandemic situations~\cite{woolliscroft2020innovation}. We must embrace ourselves by leveraging innovative technology to prepare ourselves to tackle future pandemics. Application of such innovative technologies should lead to progress, making our lives better and easier in both a physical and psychological way. Technology innovation has the potential to introduce new technologies that could support people and society during these difficult times~\cite{zeng2020high}.\\
\begin{figure}[!t]
\centering
\includegraphics[width=3.50in]{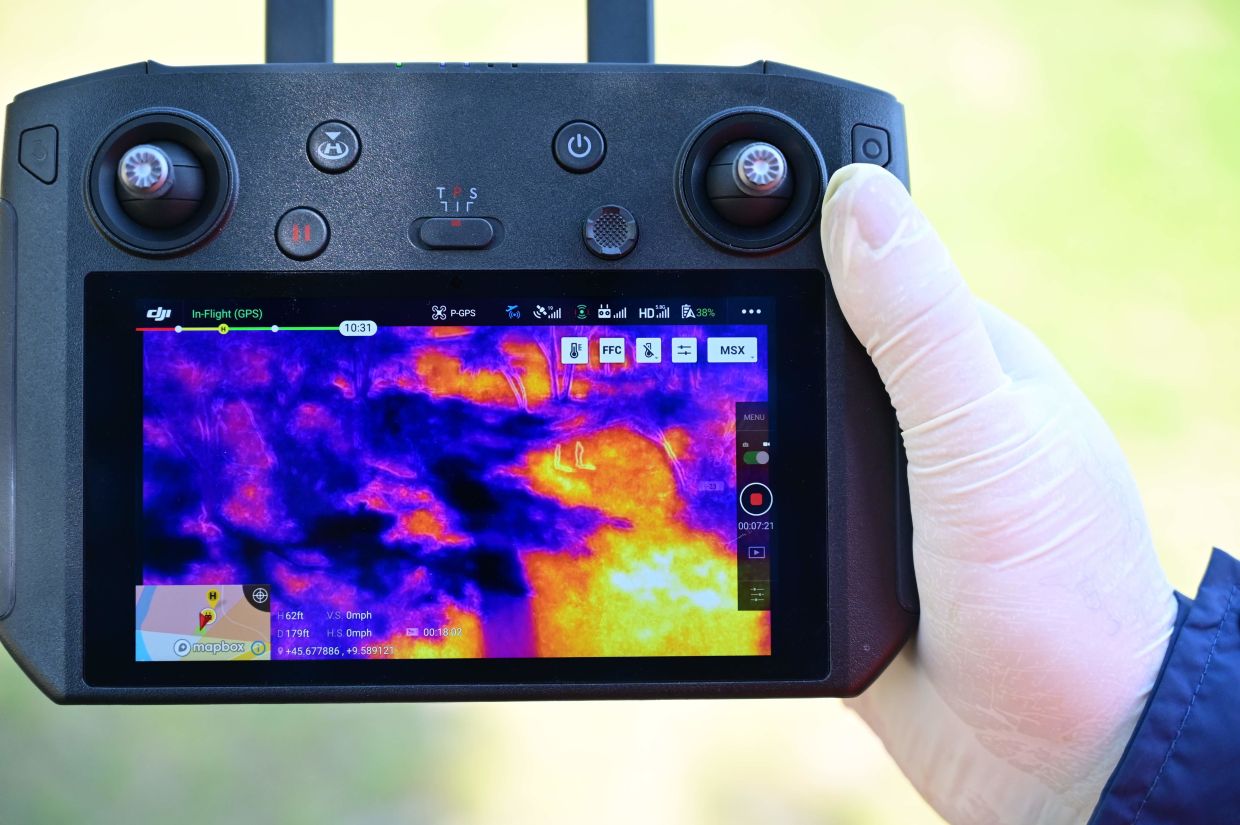}
\caption{Thermal imaging produced by a DJI Mavic 2 enterprise drone equipped with a thermal sensor for checking people's temperature deployed in Treviolo, Italy~\cite{it_2020}.}
\label{fig:figure1}
\end{figure}
\indent Nowadays, drones are commonly used for many purposes such as delivering goods, surveying, medical goods transportation, etc. These are just a few examples of using drones and solving real-life practical problems. A comprehensive review on the use of different technologies to combat COVID-19 are outlined in~\cite{chamola2020comprehensive,vargo2021digital}.

Because of the current COVID-19 situation, countries like Italy and China are already using heat-sensing drones to measure a person's body temperature~\cite{it_2020,bbc_2020}. As shown in Figure~\ref{fig:figure1}, the thermal imaging produced by drones such as DJI Mavic 2 Enterprise equipped with thermal sensors for checking people's temperature in Treviolo, Italy was already in place during the country's lockdown aimed at stopping the spread of the COVID-19 pandemic. These drones generate an alert, and a police officer acts immediately and goes to check the person's temperature. We believe that such authorities (e.g., police officers) are at risk of being infected by the reported patients. In addition to this, there will be a lot of burden on the authorities to go and check the reportedly infected persons if their sensing drones generate many false alarms. Moreover, we conjecture that the patient’s temperature can not be the only parameter used to identify such a pandemic virus. \\
\begin{figure*}[!t]
\centering
\includegraphics[width=6.0in]{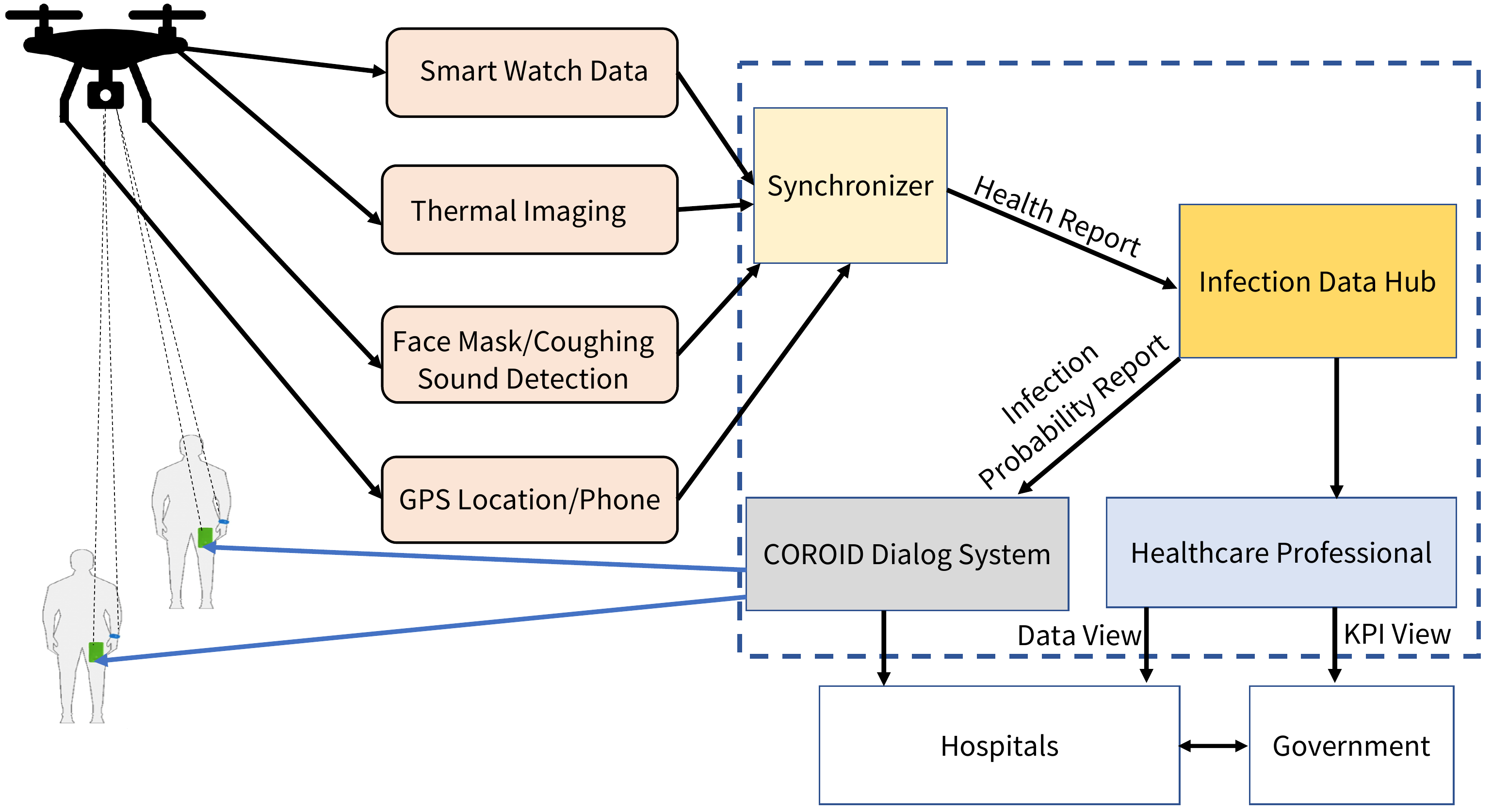}
\caption{Crowdsourcing-based system model for companion COROID drones.}
\label{fig:figure2}
\end{figure*}
\indent Recently it has been possible to collect datasets, anonymize them and monitor a large population according to the General Data Protection Regulation (GDPR)~\cite{thambawita2020pmdata}. This could be one of the promising solutions for identifying an individual with viral symptoms and monitoring large populations to address the challenges related to COVID-19 and other related future pandemics. Crowdsourcing models are believed to help correlate and do analysis with the crowdsourced real-time data to develop an action plan or decision~\cite{pan2021improving}.\\
\indent Drone technology and its application are being used in many research domains for different purposes. In the context of combating the COVID-19 pandemic, many previous studies have addressed how drone technologies can be efficiently utilized to fight the COVID-19 pandemic~\cite{restas2022drone, unicef2020drones, kumar2021drone}. However, most of these studies have not proposed any architecture that could be effectively utilized to tackle pandemic situations. There is  a potential research gap to address to propose an efficient architecture that could potentially be realized in a real-world scenario. Leveraging the idea of using drones for crowdsourcing  health-related sensors data to monitor a large population, we propose an innovative solution called COROID. COROID is a crowdsourcing-based companion drone that can tackle current and future pandemics. Our idea is novel as our companion COROID drone uses crowdsourced health-related sensors data of the person along with the readings of the cameras such as infrared camera abiding the privacy laws. \\
\indent To the best of our knowledge, this idea hasn’t been introduced either as a concept or product. Therefore, we believe our idea is novel and has huge potential to tackle current COVID-19 pandemics, and it can also be used for future pandemics. Moreover, we can use our system to monitor the patients who are supposed to be self quarantined or in isolation to prevent further the spread of the disease. In addition, we also believe that the proposed solution will be efficient and low-cost, and it could potentially be deployed in all countries.\\
\indent The rest of the paper is organized as follows.  In Section II, we explain our proposed crowdsourcing-based companion COROID drones. The System model and added features of COROID drone are outlined in Section III. Users of COROID drones are presented in Section IV. Finally, conclusions and future works are drawn in Section V.

\addtolength{\topmargin}{0.01in} 
\section{Proposed Crowdsourcing-based Companion COROID Drones}
We have named our companion drone as ``COROID" as it is designed to handle the current COVID-19 situations, but it can also be extended to tackling future pandemic situations. COROID is equipped with a camera such as an infrared camera that can be used to monitor the temperature of people in public places from a distance. With digitally collecting and using the lifelogging dataset by utilizing sensors in smart devices and gadgets, GPS location tracking has been common to monitor a person's activity to improve their health. Our COROID drone is based on the principle of using crowdsourcing data from the public, which can be used to correlate along with the reading of the infrared cameras equipped on COROID drones. Crowdsourcing enables a mobile ecosystem to collect a huge quantity of data and share different services among IoT devices~\cite{lakhdari2020composing,rauniyar2016crowdsourcing}.

Nowadays, most people use smartphones, smartwatches, and electronic gadgets to track their health conditions. Such smart devices come in varieties of flavors. Take, for example, a person wearing a smartwatch that could monitor temperature, blood pressure, heart rate, breathing rate, blood glucose level, etc. A person's smartwatch or electronic gadgets to track personal health are always connected to their smartphone. Whenever a person visits the public places where our COROID drones are already deployed in place will read the temperature of a person using the equipped infrared camera from a distance. However, it should be noted that the readings through the thermal camera are not always accurate. Our COROID drone act as an access point as it is capable of crowdsourcing people's smart wearables gadget data. COROID will correlate its infrared camera readings with the crowdsourced data from the smart devices in the connected secured cloud. In case a person's temperature is above the normal readings, and the crowdsourced data suggests abnormalities in a person, it can send an \textit{alert} to the healthcare authorities regarding the suspicion of the infected person. At the same time, it will also generate an alert to the nearby public and police authorities who are in the transmission range of COROID. Thus, COROID is believed to work as a companion for the general public and healthcare authorities to reduce the transmission of the pandemic virus. After receiving the necessary information of the susceptible users from the COROID drones, the healthcare authorities can plan out actions and act accordingly.

\vspace{-2.0ex}
\section{System Model}
\indent Our crowdsourcing-based system model for the companion COROID drone is shown in Figure~\ref{fig:figure2}. COROID drone is equipped with thermal imaging or infrared cameras. Also, these COROID drones can act as access points and sense the smartwatch/health gadgets data through a person's mobile phone connected to the COROID drone in the vicinity. The drone receives information from mobile phones using both WiFi and Bluetooth interfaces in our system model.  We collect information from persons with the COROID application installed on their mobile phones under their consent to consider privacy concerns complying with the GDPR regulations.GPS information is also accessible to COROID drones to estimate a person's location information, which then consults a mobile phone registration database of a city to have exact detail and profile of the person such as phone number, and address details. Mobile phone data has already been used for emergency management scenarios. A detailed survey on the use of mobile phone data for emergency management is outlined in~\cite{wang2020using}. The COROID drones also have the artificial intelligence capability to detect face masks and coughing sounds. Face mask detection and coughing sound detection are useful features to prevent the spread of the virus. More details on the face mask and coughing sound detection are provided in the following subsection.

The software infrastructure that integrates sensing and data processing to categorize a person's infections has four major components: the Synchronizer, the Infection Data Hub, the COROID Dialog System, and the Analytics.

\begin{itemize}
    \item \textbf{Synchronizer}. The COROID drones are installed at various locations around the city to monitor large groups of population. The raw measurement data collected by the COROID drones through cameras such as thermal imaging/infrared cameras, smartwatch/heath gadget, GPS location are synchronized by Synchronizer at each location. After synchronizing the raw data, the Synchronizer then consults a mobile registration database to obtain accurate profile information of the person. This information, along with sensors data, is subsequently sent to a centralized cloud platform from the sensing site. 
    \item \textbf{Infection Data Hub}. The Infection Data Hub in the cloud records the health information of a person, resulting in the creation of a complete registration for each person around the COROID drones. The health-related sensors data and thermal images are fused together for more robust AI-enabled classification. It is now possible to use AI on the sensors data and images together for more robust classification results~\cite{fang2021hierarchical}. The health classification results are then compared to a pre-defined infection profile. The classifications made are saved in the Infection Data Hub in the cloud.
    \item \textbf{COROID Dialog System}. Following classification, the Infection Data Hub notifies the COROID Dialog System, which transmits the infection probability result of the selected person (probably virus-infected) on his/her mobile phone and also to the city hospitals.
    \item \textbf{Healthcare Professional Analytics}. This module acts as a detailed data analysis platform for the healthcare professional to investigate health reports of the person in-depth, enriching the reasoning used to create restrictions for the probably infected person. The healthcare professional analytics platform can send the data to hospitals and key performance indicators (KPI) such as the number of probably infected persons to the local government for detailed analysis, which can be useful during the pandemic.
\end{itemize}

The following are the added features and functionalities that could be deployed on our COROID drones to monitor the crowd during the pandemics effectively.
 
\subsection{Public Mobility Monitoring}
Human mobility has demonstrated that aggregate and pseudo-anonymized mobile phone data can help to model the geographic spread of epidemics~\cite{oliver2020mobile}. The COROID drones can continuously detect and calculate human density and vehicle density as an indicator of community mobility based on pseudo-anonymized mobile phone data and camera data in public areas. It is believed that the crowded and high-density areas are at high risk of transmitting the virus during such pandemics. Suppose a designated monitoring area where our COROID drones are deployed and detect a high density of humans and vehicles. In that case, it can relay the information to the concerned authority, who can plan out the action accordingly.
\begin{figure}[!t]
\centering
\includegraphics[width=3.5in]{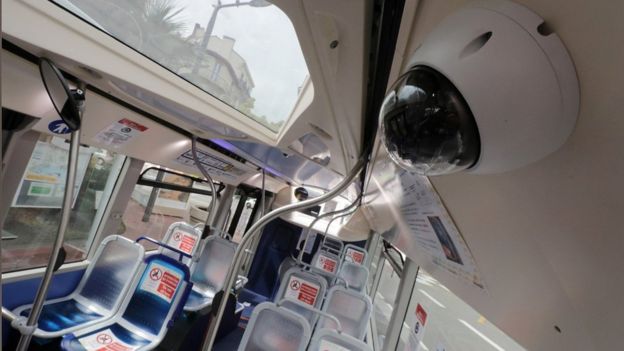}
\caption{Surveillance camera on a bus in Cannes, France~\cite{bbc}.}
\label{fig:figure3}
\end{figure}
\vspace{-1.6ex}
\begin{figure}[!t]
\centering
\includegraphics[width=3.5in]{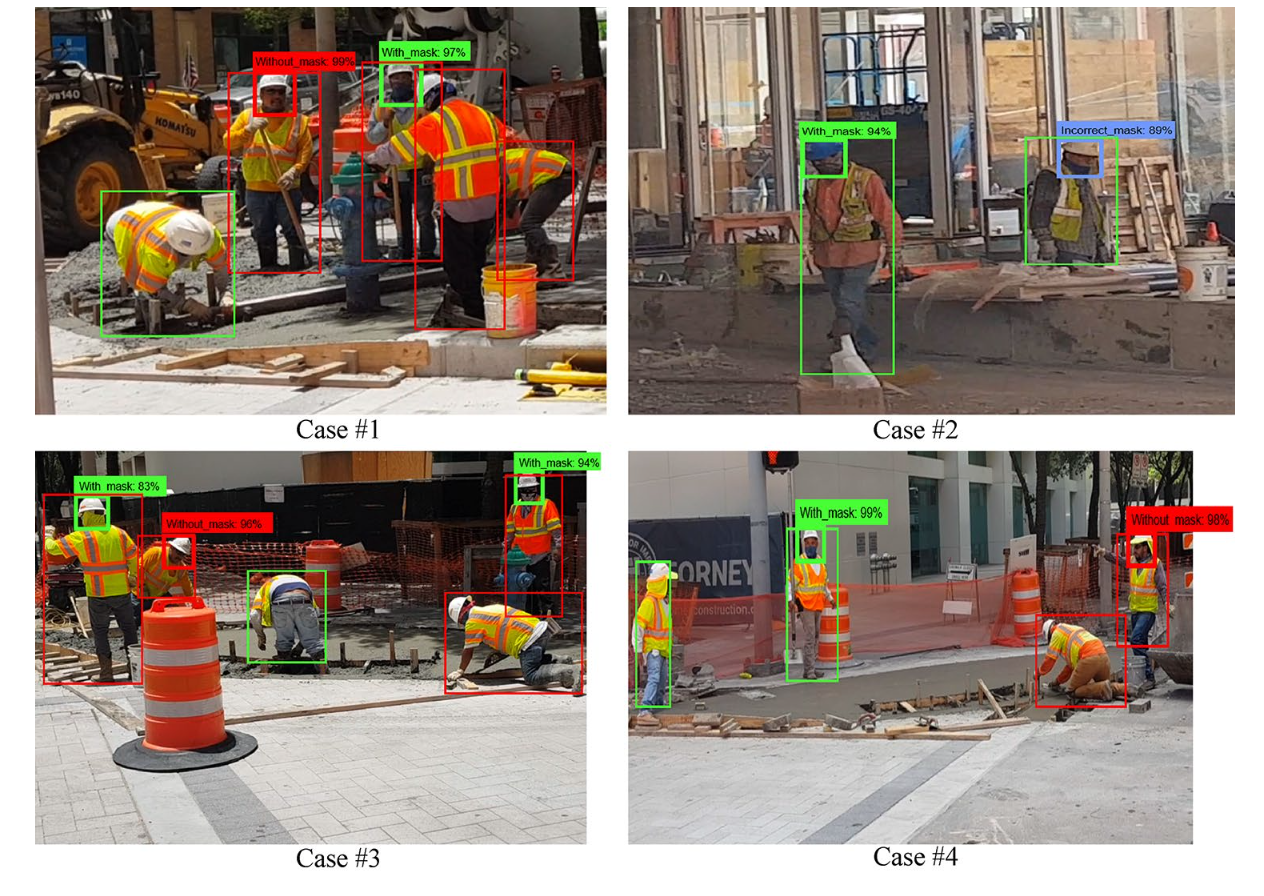}
\caption{Application of AI for face mask monitoring~\cite{razavi2022automatic}.}
\label{fig:figure4}
\end{figure}
\subsection{Face Mask Monitoring}
With COVID-19, WHO has recommended to wear face mask continuously to avoid the transmission of the virus. Countries such as France have already taken the initiative to use artificial intelligence (AI) equipped cameras to check whether people are wearing masks on public transport~\cite{vin,bbc}. An AI-equipped surveillance camera being deployed on a bus in Cannes, France, is shown in Figure~\ref{fig:figure3}. This innovative solution could send an alert if a person is not obeying the rules to wear masks in public places. With the same motive, the AI-equipped cameras on our COROID drones can automatically monitor and send a warning signal when it detects a person who is not wearing masks and violating the rules. A successful use case of monitoring construction workers for face masks is shown in Figure~\ref{fig:figure4}, which is carried out in Ref~\cite{razavi2022automatic}.

\subsection{Social Distance Monitoring}
WHO has recommended maintaining at least 1 meter (3 feet) distance with others to avoid the transmission of COVID-19~\cite{who}. It is highly possible that one may breathe in the droplets of the infected person when social distancing protocol is not followed. Therefore, many countries worldwide have made it mandatory to follow social distancing protocol~\cite{qian2020covid}. In this regard, our COROID drones can also monitor and warn automatically when detecting distances between people less than 1 meter.

\subsection{Coughing Sound Monitoring} 
Coughing sound monitoring entails collecting cough episodes from audio recordings of various real-world loud situations and then screening for COVID-19 infection based on cough features. It is now possible to use AI for cough sound monitoring to detect COVID-19 patients~\cite{zhang2021novel,orlandic2021coughvid}. To have this feature, our COROID drones can utilize its onboard microphone to capture people's coughing sound and then screen it for COVID-19 or other known infection profiles.

\indent With all these added features, we believe the capability of companion COROID drones would be significantly higher to tackle pandemic situations in the future.

\section{Users of COROID Drones}
The users of our COROID drones could be the healthcare authorities, medical personnel, police officers, government, and the general public who are concerned about being infected by any pandemic virus. Our COROID drones are a proactive solution in identifying suspected or infected people beforehand so that they can be quickly treated without transmitting the infection to others and the general public. Currently, if a person develops symptoms of any virus then they have to go to hospital or healthcare centers to get it tested for the virus infection. In the meantime, the infected person can transmit the virus to many people even without knowing they already have the virus. Therefore, our proposed solution can minimize the rates of spreading the virus and mortality rates by proactively notifying the healthcare authorities or the individuals so that the infected person can be tested at an early stage. We do not have any proactive solution to tackle such serious situations until now. Therefore, we believe that our proposed solution is practical and it can be beneficial in reducing the transmission of the infection during a pandemic.

\section{Conclusions and Future Works}
Monitoring a large crowd for viral infections during a pandemic time could be quite challenging, as observed in the COVID-19 pandemic. We discussed how innovative technologies could support people and society during these difficult times. We introduced the COROID drone, a crowdsourcing-based companion drone designed to tackle current and future pandemics. We presented the component details of the COROID drones with added features such as face mask monitoring, social distance monitoring, cough sound monitoring, etc. Hospitals can potentially utilize the infection data recorded by our COROID drones. The local government could utilize KPI values from the infection data hub for detailed analysis, such as the number of probably infected persons that can be pretty useful during the pandemic. With all these features, we believe that the proposed solution will be efficient and low-cost, and it could be potentially deployed in all the countries to tackle future pandemics.\\
\indent Field testing of the companion COROID drones to integrate all innovative features is our interest for future works. In addition to this, we will also investigate the use of AI for hierarchical multi-modal sensors and image fusion to devise enhanced classification results that could detect viral infections in real-time.

\clearpage
\bibliographystyle{IEEEtran}
\bibliography{References}
\end{document}